\pdfoutput=1
\documentclass[11pt]{article}

\usepackage{etex}
\reserveinserts{28}

\usepackage{stackrel,amsmath,amssymb,amsfonts,bbm,verbatim}

\newcommand{\leftrarrows}{\mathrel{\raise.75ex\hbox{\oalign{%
  $\scriptstyle\leftarrow$\cr
  \vrule width0pt height.5ex$\hfil\scriptstyle\relbar$\cr}}}}
\newcommand{\lrightarrows}{\mathrel{\raise.75ex\hbox{\oalign{%
  $\scriptstyle\relbar$\hfil\cr
  $\scriptstyle\vrule width0pt height.5ex\smash\rightarrow$\cr}}}}
\newcommand{\Rrelbar}{\mathrel{\raise.75ex\hbox{\oalign{%
  $\scriptstyle\relbar$\cr
  \vrule width0pt height.5ex$\scriptstyle\relbar$}}}}

\makeatletter
\def\leftrightarrowsfill@{\arrowfill@\leftrarrows\Rrelbar\lrightarrows}
\newcommand{\xleftrightarrows}[2][]{\ext@arrow 3399\leftrightarrowsfill@{#1}{#2}}
\makeatother

\usepackage[multiple,stable]{footmisc}
\usepackage[amsmath,amsthm,thmmarks]{ntheorem}
\allowdisplaybreaks[4]
\usepackage[noconfig]{refstyle}
\usepackage[dvipsnames]{xcolor}
\usepackage[hyperfootnotes=false, linktocpage=true, colorlinks, citecolor=blue, linkcolor=blue, urlcolor=Maroon]{hyperref}
\usepackage{array,tabularx,booktabs,geometry,subfigure,graphicx,longtable}
\newcolumntype{M}[1]{>{\centering\arraybackslash}m{#1}}
\newcolumntype{P}[1]{>{\centering\arraybackslash}p{#1}}

\usepackage[bf]{caption}
\usepackage{tikz}
\usetikzlibrary{shapes,arrows}
\tikzstyle{block} = [rectangle, draw, text width=7em, text centered, rounded corners, minimum height=3em]
\usepackage{graphicx,color}
\usetikzlibrary{positioning,arrows}

\usepackage{cite}
\usepackage{arydshln}
\usepackage{tikz}
\usetikzlibrary{positioning,automata,decorations.pathreplacing,shapes}
\usepackage[english]{babel}
\usepackage{microtype}

\usepackage[all,cmtip]{xy}
\usetikzlibrary{matrix, arrows}

\numberwithin{equation}{section}

\usepackage{hyperref}
\usepackage{extarrows}

\newcommand{\eeq}{\end{equation}}
\newcommand{\beq}{\begin{equation}}
\newcommand{\ba}{\begin{array}}
\newcommand{\ea}{\end{array}}

\newcommand{\cV}{{\cal V}}

\newcommand{\IP}{\mathbb P}

\def\IP{\mathbb{P}}

\def\IZ{\mathbb{Z}}

\newcommand{\id}{\mathbf{1}}

\def\cP{\mathcal{P}}

\def\cV{\mathcal{V}}

\def\clap#1{\hbox to 0pt{\hss#1\hss}}

\newcommand{\be}{\begin{equation}}
\newcommand{\ee}{\end{equation}}
\newcommand{\eea}{\end{eqnarray}}
\newcommand{\bea}{\begin{eqnarray}}

\newcommand{\iddots}{\mathinner{\mkern2mu\raise1pt\hbox{.}\mkern2mu \raise4pt\hbox{.}\mkern2mu\raise7pt\hbox{.}\mkern1mu}}

\providecommand{\id}{\leavevmode\hbox{\small$\mathrm{1}$\kern-3.8pt\normalsize$\mathrm{1}$}}

 \newcommand{\bi}{\begin{itemize}}
\newcommand{\ei}{\end{itemize}}

\begin{document}

\begin{flushright}
{\tt\normalsize CTPU-PTC-21-42}
\end{flushright}

\vskip 40 pt
\begin{center}
{\Large \bf 
On Towers of Light States at Infinite Distance
}

\vskip 10 mm

Seung-Joo Lee${}^{}$

\vskip 5 mm
\small ${}^{}${\it Center for Theoretical Physics of the Universe, \\ Institute for Basic Science, Daejeon 34126, South Korea} \\[3 mm]

\vskip 3mm
Based on a talk at the ``Nankai Symposium on Mathematical Dialogues,'' 2021.

\end{center}

\vskip 5mm

\begin{abstract} 
\vskip 1mm

Upon investigating asymptotic regimes of the F-theory moduli space, we verify that a tower of light states arises as predicted by the Distance Conjecture. Specifically, we provide a geometric classification of the infinite distance limits and comprehensively analyze the light states of the associated effective theories. We thereby find for every infinite distance limit that the effective theory either reduces to a weakly-coupled (dual) string theory or decompactifies to a higher-dimensional theory. This is in full agreement with the Emergent String Conjecture, which clarifies the physical nature of the light particle tower either as the excitation modes of an emergent weakly-coupled string or as the Kaluza-Klein modes associated with a decompactification of the spacetime. The results reported encompass both the K\"ahler and the complex structure limits of F-theory, respectively, in $4$ and $8$ dimensions, and hence are indicative of the overarching microscopic origins of the light towers. 

\end{abstract}

\vskip 5mm

\section{Introduction}
The notion of the {\it Landscape}, as the set of all consistent low-energy effective theories of quantum gravity, has served for decades as a central idea behind much research activity in high energy physics. Perhaps an equally important notion is that of the {\it Swampland}, which is the complementary set to the Landscape, hence collecting all the effective theories that cannot incorporate quantum gravity at high energies. 
Whether it is for intellectual guidance towards new physics ideas in model building, or for deeper understanding of the mysterious nature of quantum gravity {\it per se}, it is important to come up with concrete criteria by which theories in the landscape can be distinguished from the ones in the swampland. This is one of the major goals of the so-called Swampland Program~\cite{Vafa:2005ui}. In this context, many general properties have thus far been proposed as universal features that every consistent theory in the landscape is believed to exhibit, which go under the name of {\it swampland conjectures}. 

Interestingly, while the individual conjectures have been put forward based on their own physical rationale, accumulating pieces of evidence continue to be revealed for the exciting interconnections amongst them, leading to an intricate web of conjectures. Arguably at the heart of this web lies the Swampland Distance Conjecture~\cite{Ooguri:2006in}, which asserts that a tower of states is bound to become exponentially light at infinite distance in moduli space of quantum gravity; it is these light states that oftentimes serve in turn as the physical objects of which presence is predicted by other conjectures,\footnote{We refer the readers e.g. to the reviews~\cite{Brennan:2017rbf, Palti:2019pca, vanBeest:2021lhn,Grana:2021zvf} for general overview of the swampland program as well as for detailed account of many important and exciting conjectures.} thereby providing fruitful insights and inspirations for the latter.\footnote{For an illustration, we briefly describe at the end of Section~\ref{Kahler} how this light tower can be argued to include the Weak Gravity sublattice/tower and to occupy the full charge lattice.}  

Despite the distinguished role played by the Swampland Distance Conjecture, the physical nature and the microscopic, dynamical origin of the predicted light states remain elusive. 
Provided that string theory is an ideal framework to investigate the quantum nature of gravity in a controlled and computable setup, one can thus hope to better understand and gain intuitions on the conjecture by clarifying whether and how it is realized in string theory. In this article we therefore address its string theoretic aspect by invoking the recent proposal of the Emergent String Conjecture~\cite{Lee:2019wij}, which identifies the light tower of states either as the excitation modes of a light string (emergent string phase) or as the light Kaluza-Klein (KK) modes (decompactification phase). 

The aim of this article is to explain how these two phases are universally realized by the asymptotic forms of the compactification geometry in the general setting of F-theory~\cite{Vafa:1996xn}, which provides a non-perturbative description of Type IIB string theory in presence of $7$-branes. To this end we will first classify infinite distance limits in the moduli space of elliptic Calabi-Yau $(n+1)$-folds $X_{n+1}$,  
\beq\label{pi}
\pi: X_{n+1} \to B_n \,,
\eeq
over the Type IIB internal $n$-folds $B_n$. 
Based on such a geometric classification, we will then proceed to characterize the associated $(10-2n)$-dimensional effective theories in the boundary of moduli space. 
Specifically, in Sections~\ref{Kahler} and~\ref{Complex}, we will discuss the geometry and the physics at infinite distance in the K\"ahler and the complex structure moduli of F-theory, respectively, in $4$ and $8$ dimensions, {\it i.e.}, with $n=3$ and $n=1$ in~\eqref{pi}; the reported results can thus be viewed as providing intuitions on most general asymptotic regimes of F-theory moduli space.\footnote{For recent progress towards the physics of infinite distance limits in various perturbative string setups, see e.g.~\cite{Corvilain:2018lgw, Baume:2019sry} and~\cite{Blumenhagen:2018nts, Grimm:2018ohb, Joshi:2019nzi, Grimm:2019ixq, Gendler:2020dfp, Grimm:2020cda, Grimm:2021ikg, Bastian:2021eom, Palti:2021ubp, Bastian:2021hpc,Grimm:2021ckh}, which, respectively, analyze the K\"ahler and the complex structure limits. See also~\cite{Xu:2020nlh} for related discussions on the M-theoretic front.} We will conclude in Section~\ref{Conc} with summary and outlook. 


\section{K\"ahler Deformations}\label{Kahler}
While the systematic study of the K\"ahler limits in F-theory originally started in a $6d$ setting~\cite{Lee:2018urn, Lee:2018spm}, in this section we will discuss more general results obtained for $4d$ $N=1$ F-theory, as analyzed more recently in~\cite{Lee:2019tst, Klaewer:2020lfg}.  
The K\"ahler moduli of $4d$ F-theory control the volumes of various cycles in the Type IIB $3$-fold $B_3$, and hence, the strengths of the couplings in the effective theory in particular. At the same time the K\"ahler geometry also governs many other physical quantities, most importantly, the tensions of various solitonic strings arising from the D3 branes wrapped on complex curves. An intriguing link may thus arise between the coupling constants and the string tensions via the common K\"ahler geometry, resulting, e.g., in a stringy realization of the Weak Gravity Conjecture in the weak gauge coupling regime. We will come back to this aspect towards the end of the section after presenting more general results applicable to all the K\"ahler limits.

We start by noting that a most general K\"ahler deformation can be thought of as the combination of an overall and a relative scalings. Infinite distance limits are therefore naturally characterized by a pair of parameters, $\mu$ and $\lambda$, which are individually taken to infinity for the asymptotic scalings of the two kinds,
\beq\label{2lims}
\mu \to \infty \quad \text{(overall scaling limit)} \,; \qquad 
\lambda \to \infty \quad \text{(relative scaling limit)} \,.
\eeq
Firstly, the former, overall limit is responsible for the linear scaling of the K\"ahler class $\mathcal J_{B_3}$,
\beq
\mathcal J_{B_3} \sim \mu J_{B_3} \,. 
\eeq
Here, by definition, the rescaled K\"ahler class $J_{B_3}$ leads to a finite rescaled volume of $B_3$, i.e.,
\beq\label{Vfin}
V_{B_3} = \frac{1}{3!}\int_{B_3} J_{B_3}^3 \sim 1\,,
\eeq
while the actual volume of $B_3$ diverges in the limit $\mu \to \infty$ as
\beq
\mathcal V_{B_3} = \frac{1}{3!} \int_{B_3} \mathcal J_{B_3}^3 \sim \mu^3 \to \infty \,. 
\eeq
On the other hand, the relative limit governs the growth of the largest K\"ahler parameter appearing in the rescaled K\"ahler class $J_{B_3}$,
\beq
J_{B_3} = \lambda J_0 + \sum_{i} t^i J_i \,. 
\eeq
Here, the classes $J_0$ and $J_i$ are the K\"ahler cone generators of $B_3$ and the associated parameters $\lambda=:t^0$ and $t^i$ are subject to the parametric relations, 
\beq
t^i \ll  t^0 \quad \text{or} \quad t^i \sim t^0 \,, \qquad \forall i\,,
\eeq
because $t^i < \lambda$ by definition of the large parameter $\lambda$. 
Combined with the finiteness of the rescaled volume~\eqref{Vfin}, the limit $\lambda \to \infty$ thus enforces vanishing of some $t^i$, and hence, (relative) shrinking of certain holomorphic curves in turn. 

\begin{figure}[t]
  \center
  \includegraphics[width=0.7\textwidth]{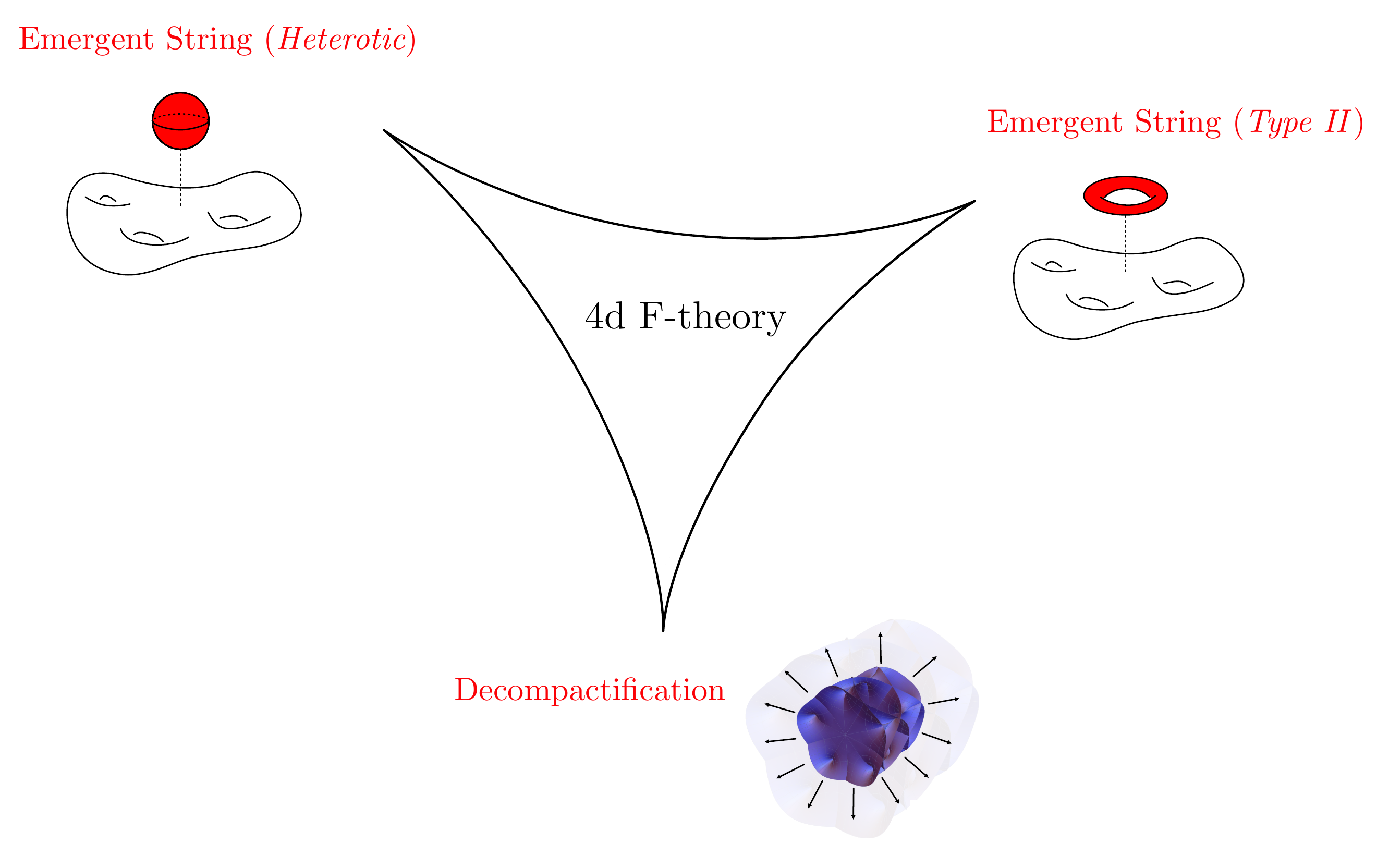}
  \caption{\it Classification of infinite distance limits in the K\"ahler moduli space of $4d$ F-theory. Depicted in the top-left and the top-right corners are the limits corresponding to an emergent string phase, governed by the asymptotically weakly-coupled heterotic and Type II strings, respectively, and in the bottom corner are those corresponding to a decompactification phase. }
  \label{fig:Kahler}
\end{figure}

As it turns out, depending largely on topology and geometry of such shrinking curves, we can classify the infinite distance K\"ahler limits into three qualitatively distinct classes as follows~\cite{Klaewer:2020lfg} (see Fig.~\ref{fig:Kahler} for a pictorial summary). 

\begin{enumerate}
\item  $B_3$ admits a $\mathbb P^1$-fibration of which fiber shrinks at a rate parametrically faster than any other non-contractable curves. D3 brane wrapping this fiber leads to an asymptotically tensionless {\it heterotic} string. 
\item  $B_3$ admits a $T^2$-fibration of which fiber shrinks at a rate parametrically faster than any other non-contractable curves. D3 brane wrapping this fiber leads to an asymptotitcally tensionless {\it Type II} string. 
\item  No unique $\mathbb P^1$ or $T^2$ fiber shrinks at a parametrically fastest rate (and hence, complementary to the previous two classes). However, expanding curves and surfaces have KK scale that wins against the tension of the solitonic critical strings, if any, indicating a decompactification. 
\end{enumerate}

\noindent
For a more detailed description of the above three classes, let us denote by $C_0 \in H_2(B_3, \mathbb Z)$ the homology class of the described fastest-shrinking fiber with genus $g(C_0) = 0$ or $1$, and similarly, by $C_\infty \in H_2(B_3, \mathbb Z)$ and $S_\infty \in H_4(B_3, \mathbb Z)$ the homology classes of non-contractable curves and surfaces, respectively, that expand at a fastest parametric rate. 

In the first two classes, the tension $M_{\rm str}$ of the relevant critical string turns out to scale as
\beq
\frac{M_{\rm str}^2}{M_{\rm Pl}^2} \sim \frac{\cV_{C_0}}{\cV_{B_3}}  \sim \mu^{-2} \lambda^{-2} \,,
\eeq
and the leading KK scale $M_{\rm KK}$ as 
\beq
\frac{M_{\rm KK}^2}{M_{\rm Pl}^2} \sim {\rm min} \left( \frac{\cV_{C_\infty}^{-1}}{\cV_{B_3}}, \frac{\cV_{S_\infty}^{-1/2}}{\cV_{B_3}}, \frac{\cV_{B_3}^{-1/3}}{\cV_{B_3}} \right) \sim \mu^{-4} \lambda^{-1} \,.
\eeq
The two scales thus compare as
\beq
\frac{M_{\rm str}^2}{M_{\rm KK}^2} \sim \mu^2 \lambda^{-1} \,,
\eeq
according to which we can naturally subdivide the space of large parameters $\lambda$ and $\mu$ into three regimes, each exhibiting a characteristic phase of the effective physics as follows: 
\bea
&\lambda \ll \mu^2& \quad \longleftrightarrow \qquad  \text{decompactification} \\ 
&\lambda \sim \mu^2& \quad \longleftrightarrow  \qquad \text{emergent string}  \\  \label{pathology}
&\lambda \gg \mu^2&  \quad \longleftrightarrow \qquad \text{potential pathology} 
\eea
Note that the third regime~\eqref{pathology}  is described as being potentially pathological since the KK tower is decoupled from the string tower, contradicting that the critical dimension for a superstring is $10$. In the first class of limits where the relevant light string is heterotic, it turns out that the day is saved by quantum corrections which are generically present with only $4$ real supercharges constraining the effective theory: careful investigation of both perturbative $\alpha'$ and non-perturbative corrections teaches us that the limits in the regime~\eqref{pathology} is obstructed. In the second class where the relevant light string is Type II, on the other hand, T-duality turns out to save the day.

In the last remaining class of limits, the fastest-shrinking fiber class $C_0$ is not uniquely defined. This indicates the presence of multiple critical strings which are equally relevant for physics. As already stated, however, the KK scale turns out to always win against the string tension, evading the pathology by entering into a decompactification phase instead. 


Finally, let us briefly discuss how the addressed tower of light states at infinite distance can connect to other swampland conjectures, most prominently to the Weak Gravity Conjecture~\cite{Arkani-Hamed:2006emk} in its strongest, sublattice version~\cite{Cheung:2014vva, Heidenreich:2015nta, Heidenreich:2016aqi, Montero:2016tif}. For a quantitative control, we restrict our discussions to a special type of infinite distance limits where the $U(1)$ vector multiplets are asymptotically weakly coupled. As it turns out, of the three classes of limits discussed so far in full generality, only the first can arise, and hence, the heterotic string is the relevant asymptotically tensionless critical string. Then, not only do we know that there will arise a light tower but we can also be more precise about the spectrum. Specifically, a subset of the full spectrum is captured by the elliptic genus of the $N=(0,2)$ heterotic worldsheet theory, of which (quasi-)Jacobi nature significantly constrains its functional behavior~\cite{Lee:2020gvu, Lee:2020blx}. This enables us to identify a sublattice\footnote{The index for the embedding of this sublattice into the full charge lattice is determined as the intersection of the shrinking curve $C_0$ with the expanding gauge divisor $S_{\infty}$ in the Type IIB internal space.}  amount of charged particles for which the Coulomb repulsion is marginally stronger than the combined attraction of the gravitational and the Yukawa forces, 
\beq\label{WGC}
|F_{\rm Coulomb}| \geq |F_{\rm Grav}| +|F_{\rm Yuk}| \,,
\eeq
as predicted by the Sublattice Weak Gravity Conjecture in presence of scalar fields~\cite{Palti:2017elp}. One can in fact verify that not only do these sublattice particles obey~\eqref{WGC} at the classical level~\cite{Lee:2019tst}, but also upon incorporating quantum corrections~\cite{Klaewer:2020lfg}.


\section{Complex Structure Deformations}\label{Complex}
Having observed that the infinite distance limits in the K\"ahler moduli space always lead to a light particle tower of either string or KK excitation modes, we now change tack and consider the limits in the complex structure moduli space, which, together with the K\"ahler counterpart, comprise all deformations of F-theory. The complex structure deformations of F-theory encompass in particular the brane moduli in the open-string sector, which has largely been ignored until recently in study of infinite distance limits. In this section we will thus focus on the infinite complex structure limits of F-theory, in an arguably simplest non-trivial context of $8d$ effective theories. 

To this end we investigate the complex structure deformations of elliptic K3 surfaces $X_u$, where $u \in D := \{ u \in \mathbb C: |u| < 1\}$ is the deformation parameter. Viewed as a fibration over the unit disk $D$, the degeneration is thus described by a $3$-fold family $\mathcal X$, whose generic member $X_{u \neq 0}$ is a smooth K3 surface while the central element $X_{u=0}$ degenerates into component surfaces  
\beq\label{X0}
X_0 = \bigcup_{p=0}^P \bigcup_{i_p=1}^{N_p} X^p_{i_p} \quad \longrightarrow \quad B_0 = \bigcup_{p=0}^P B^p \,,
\eeq
where $X^p_{i_p}$ are fibered over the rational base components $B^p$. 
The family $\mathcal X$ is called a Kulikov model only if a set of extra conditions are satisfied.\footnote{A degeneration is semi-stable if the components of the degenerate element $X_0$ are reduced and $X_0$ only involves normal-crossing singularities. Kulikov model is in turn defined as a semi-stable degeneration of K3 surfaces $X_u$ of which $3$-fold family $\mathcal X$ is Ricci-flat.} However, it is known that any degeneration can be put into such a Kulikov form by an appropriate combination of base changes and birational transformations~\cite{Kulikov1, Kulikov2, PerssonPink}, which, importantly, do not affect the effective physics.\footnote{Base change is a map $u\to u^k$ for $k\in \mathbb Z$ and hence does not alter the surface defined precisely at $u=0$. The relevant birational transformations turn out to be base blowups, which correspond to zooming in towards the brane collision.} Without loss of generality we may thus restrict only to the Kulikov models in analyzing the physics of complex structure degenerations. 

Kulikov models are classified into Type I, II and III, where the latter two are known to be at infinite distance~\cite{Kulikov2, FriedmanMorrison, Persson}. Our focus in this section thus lies in Kulikov models of Type II and III. Amongst many established properties of theirs, we point out that the surface components $X^p_{i_p}$ intersect at an elliptic and a rational curve, respectively, in Type II and III Kulikov models. Moreover, Type II models exhibit a pair of shrinking transcendental 2-tori, $(\gamma_1, \gamma_2)$, and Type III models a single such 2-torus, $\gamma_1$. As in the K\"ahler limits of the previous sections, these 2-tori, due to their shirking nature in the complex structure limits, turn out to play a pivotal role in generating a light particle tower, which we will address momentarily. 

\begin{figure}[t]
  \center
  \includegraphics[width=0.6\textwidth]{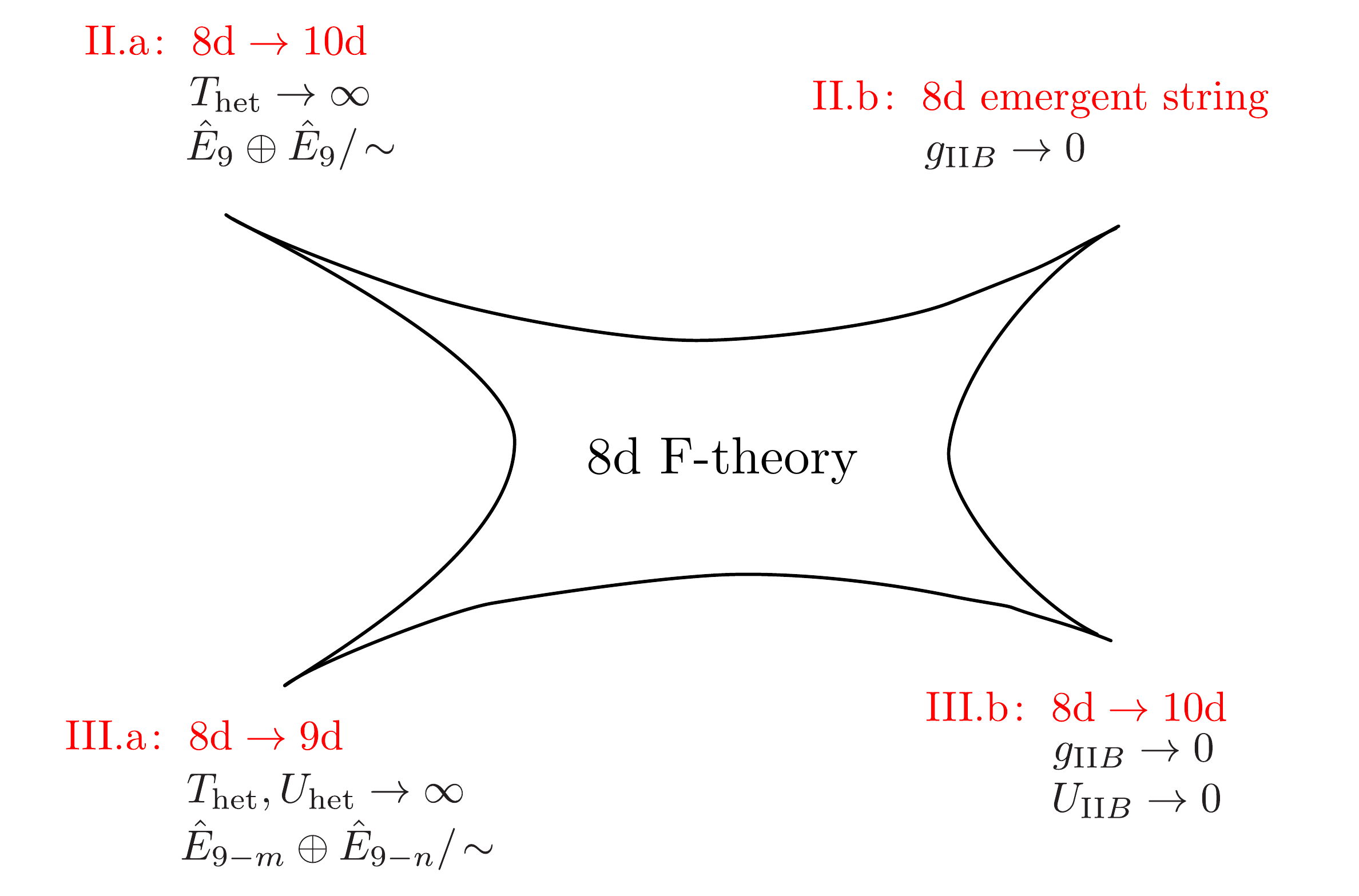}
  \caption{\it Classification of infinite distance limits in the complex structure moduli space of $8d$ F-theory via refinement of elliptic Kulikov models. The limits in the top-right corner (models of Type II.b) correspond to an emergent string phase and those in the other three corners lead to a decompactification phase either to $9d$ (Type III.a) or to $10d$ (Type II.a and III.b). }
  \label{fig:CS}
\end{figure}

Importantly, it turns out that elliptic Kulikov models of Type II and III can each be refined into two subtypes, so that the effective physics associated to each of the total of four subtypes exhibits a distinguished feature. In this context, the results that we will describe in this section are summarized in Fig.~\ref{fig:CS}. 
To begin with, the refinement of elliptic Type II Kulikov models into Type II.a and II.b was already established in~\cite{Clingher:2003ui} as follows.\footnote{The described canonical forms of $X_0$ can be achieved by appropriate base changes and birational transformations.}   
\bi
\item {\bf II.a:} 
The central element $X_0$ takes the form~\eqref{X0} with $P=1$ and $N_0=N_1=1$: the two components $X^0_1$ and $X^1_1$ are rational elliptic surfaces and meet along the common smooth elliptic fiber sitting at $B^0 \cap B^1$ in the base.  

\item {\bf II.b:} The central element $X_0$ takes the form~\eqref{X0} with $P=0$ and $N_0=2$: the two components $X^0_1$ and $X^0_2$ are $\IP^1$-fibrations over the common rational base $B^0$ and meet along its double cover ramified over four base points, which is thus an elliptic curve. 

\begin{figure}[h!]
  \center
  \includegraphics[width=0.9\textwidth]{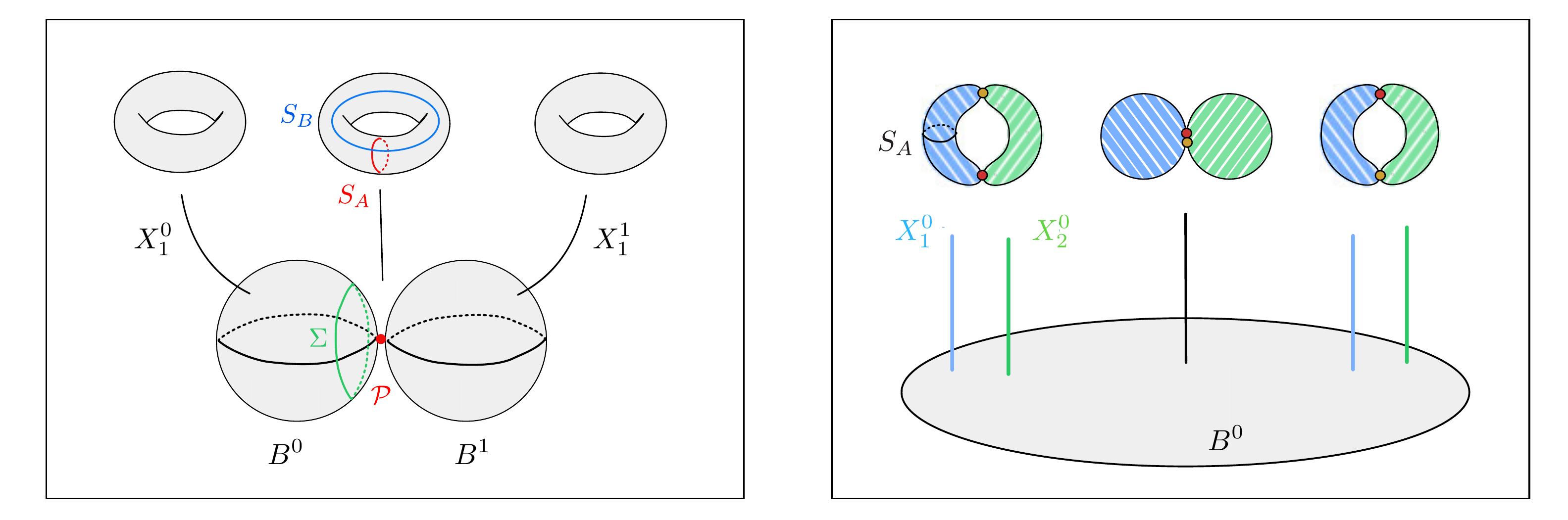}
  \caption{\it Refinement of elliptic Kulikov models of Type II. Models of Type II.a (left) exhibit a degeneration of K3 surface into two rational elliptic components $X^0_1$ and $X^1_1$ meeting along the common smooth elliptic fiber, which sits at the intersection $\cP$ of their bases $B^0$ and $B^1$. In models of Type II.b (right) the K3 degenerates into $X^0_1$ and $X^0_2$, each of which is a $\IP^1$-fibration over the common base $B^0$; these two components meet along the elliptic curve realized as the double cover of $B^0$. }
  \label{fig:TypeIIab}
\end{figure}

\ei
See Fig.~\ref{fig:TypeIIab} for an illustration of Type II.a and II.b Kulikov degenerations. 

The physics of the associated effective theory had also been well-known. F-theory on a Type II.a Kulikov model is dual to heterotic string on a torus with the K\"ahler parameter taken to infinity ($T_{\rm het} \to \infty$)~\cite{Morrison:1996na, Morrison:1996pp, Aspinwall:1997ye}, and hence, the spacetime fully decompactifies to $10d$. On the other hand a Type II.b model realizes the so-called Sen's limit~\cite{Sen:1996vd, Aspinwall:1997ye, Clingher:2012rg}, where the effective theory is described as a weakly-coupled Type IIB string in $8d$. 
In fact, the characteristic structure for the geometry of these two subtypes enables us to immediately identify the described phases of the effective physics. 
\bi
\item For models of  Type II.a, decompactification of two circular directions is indicated in M-theory by the pair of light KK-like towers, which result from M2 branes wrapping the shrinking transcendental 2-tori, $(\gamma_{1}, \gamma_2)$. The latter in turn arise from fibering the A- and the B-cycles of the elliptic fiber $X^0_1 \cap X^1_1$ along the vanishing 1-cycle encircling the point $B^0\cap B^1$ in the base; see Fig.~\ref{fig:TypeIIab} (left). These states correspond in F-theory to the towers resulting from $(1,0)$- and $(0,1)$-strings wound around the same vanishing 1-cycle of the base. 
\item For models of Type II.b, 
the ``elliptic'' fiber over the base $B^0$ has its A-cycle shrinking, and hence, the shirking 2-tori, $(\gamma_1, \gamma_2)$, arise from adjoining this vanishing A-cycle in the fiber over the two 1-cycles in the base, to be precise, in the double-cover elliptic curve $X^0_1 \cap X^0_2$; see Fig.~\ref{fig:TypeIIab} (right). M2 branes wrapping those 2-tori again lead to a pair of light KK-like towers. However, there also arises a tensionless $(1,0)$-string from M2 brane wrapping only the vanishing A-cycle. As it turns out, the KK and the string scales are parametrically the same, and hence, the spacetime does {\it not} decompactify.
\ei

More recently, elliptic Kulikov models of Type III have also been refined in the same vein into Type III.a and III.b models~\cite{Lee:2021qkx}.\footnote{See also the related independent analysis of~\cite{alexeev2021compactifications, Brunyantethesis, Odaka:2018kua,Odaka:2020ezm, ascher2021compact} regarding compactification of the moduli space of elliptic K3s.} For a concise description of this refinement, let us first consider the {\it Kulikov Weierstrass model} $\mathcal Y$ by blowing down an elliptic Kulikov model $\mathcal X$ along the fiber, i.e., by contracting all the exceptional fibral curves. The central element $Y_0$ of $\mathcal Y$ thus decomposes as
\beq\label{Ydecomp}
Y_0 = \bigcup_{p=0}^P Y^p  \quad \longrightarrow \quad B_0 = \bigcup_{p=0}^P B^p \,,  
\eeq
of which base $B_0$ is unchanged from~\eqref{X0}. 
One can then blow $\mathcal Y$ further down along the base to an even more degenerate model $\widehat{\cal Y}$ by contracting all but one of the base components $B^p$. The central element $\hat Y_0$ of $\widehat{\mathcal Y}$ would then consist of a single surface component, elliptically fibered over a rational curve base $\hat B_0$, 
\beq\label{hatY0}
\hat Y_0 \quad \longrightarrow \quad \hat B_0 =: \mathbb P^1_{[s:t]} \,, 
\eeq
where $[s:t]$ are the homogeneous coordinates of the base. In terms of the Weierstrass description of the F-theory K3 surface, 
\beq \label{Weierstrassfam-def}
\hat Y_u: \qquad y^2 = x^3 + f(s,t; \,u) x z^4 + g(s,t; \,u) z^6 \,, 
\eeq
it is in fact most natural to view the infinite complex structure limit as being initially realized in $\hat Y_0$. 
The involved birational transformations are schematically summarized as
\beq
\mathcal X
\xleftrightarrows[\text{fibral blowdown}]{\text{fibral blowup}} \mathcal Y
\xleftrightarrows[\text{base blowdown}]{\text{base blowup}} \widehat{\mathcal Y}  \,.
\eeq

As is well-known, the codimension-one singular fibers of a Weierstrass model $\hat Y_0$ can be characterized by the vanishing orders of the sections $f$ and $g$, as well as the discriminant $\Delta:=4f^3 + 27 g^2$, via the Kodaira-N\'eron classification. Here, the so-called {\it minimal} fiber in codimension-one locus of the base $\hat B_0=\IP^1_{[s:t]}$, e.g. at $s=0$, is subject to  
\beq
{\rm ord}_{\hat Y_0} (f)|_{s=0} <4 \quad \text{or} \quad {\rm ord}_{\hat Y_0} (g)|_{s=0} <6 \,,
\eeq
and represents a finite non-Abelian enhancement of the gauge algebra of the effective theory. Potential infinite distance limits must therefore exhibit either a codimension-one {\it non-minimal} fiber with  
\beq
{\rm ord}_{\hat Y_0} (f, g, \Delta)|_{s=0} = (4+\alpha, 6+\beta, 12+\gamma) \,,\quad \text{for}\quad \alpha, \beta, \gamma \geq 0 \,,
\eeq
or else generic singular fibers in {\it codimension zero} (in absence of non-minimal fibers in codimension one). It has indeed been shown~\cite{Lee:2021qkx} that the former can lead to degenerations of Type II.a (if $\gamma=0$) or of Type III (if $\gamma > 0$),\footnote{To be precise, $\gamma>0$ together with $\alpha=\beta=0$ guarantee that the associated Kulikov model is of Type III. If $\alpha$, $\beta$, and $\gamma$ are all strictly positive, even a Type I Kulikov model might arise, as exemplified in~\cite{Lee:2021qkx}.} and the latter, to those of Type II.b.

\begin{figure}[t]
  \center
  \includegraphics[width=1\textwidth]{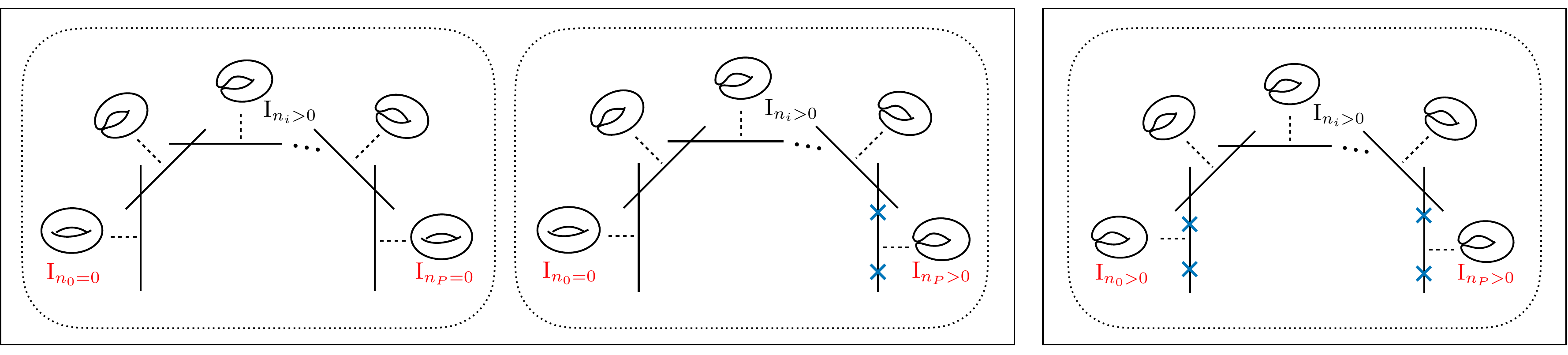}
  \caption{\it Refinement of elliptic Kulikov models of Type III in terms of their Weierstrass models $\mathcal Y$. Depicted are the characterization of possible configurations for the central element $Y_0$ of $\mathcal Y$. The degenerate surface $Y_0$ decomposes into a chain of components $Y^p$ with generic fibers of Kodaira types I$_{n_p}$ for $0\leq p \leq P$, with $P \geq 2$. The middle components $Y^p$ for $0<p<P$ have generic fibers of Kodaira type I$_{n_p}$ with $n_p >0$. Elliptic Type III models are thus refined into Type III.a models where none/one of the end components degenerates, respectively, as in the first/second configuration (left), and Type III.b models where both ends degenerate as in the third configuration (right). Blue crosses on the degenerate end components denote D-type fibers.}
  \label{fig:TypeIIIab}
\end{figure}

As it turns out, however, the refinement of elliptic Type III Kulikov models is more naturally seen from the central element $Y_0$ (as opposed to $\hat Y_0$), of which decomposition~\eqref{Ydecomp} can be proven to take a chain form. Furthermore, $P \geq 2$ can be assumed without loss of generalities,\footnote{This is because $P\geq 2$ can always be achieved by a base change.} leading to a natural division of the components $Y^p \to B^p$ into the middle and the end surfaces, respectively, for $0<p<P$ and for $p=0$ or $P$. The refined classification is then motivated by the types of the generic fibers over the base components $B^p$, which are necessarily of Kodaira types I$_{n_p}$ for $\cal X$ to be a Kulikov model.\footnote{Note that $n_p$ coincide with $N_p$ in~\eqref{X0} if the generic fibers are singular, i.e., if $n_p>0$.} One can then show for any Type III Kulikov Weierstrass models that $n_p >0$ for the middle components, and hence, elliptic Type III models are naturally refined by the degeneration types of the end components $Y^0$ and $Y^P$ as follows~\cite{Lee:2021qkx}.   
\bi
\item {\bf III.a:} The generic fibers of at least one end component are smooth, i.e., $n_0=0$ or $n_P=0$.       
\item {\bf III.b:} The generic fibers of both end components are singular, i.e., $n_0 >0$ and $n_P >0$. 
\ei
Furthermore, special fibers of the degenerate components with $n_p>0$ are restricted to only take one of the two forms,   
\beq\label{AD}
\text{A-type}:~ {\rm ord}_{Y^p} (f,g,\Delta)|_{{\rm pt}\in B^p} = (0,0,k) \quad \text{or} \quad  \text{D-type}:~ {\rm ord}_{Y^p} (f,g,\Delta)|_{{\rm pt}\in B^p} = (2,3,k) \,, 
\eeq
where D-type fibers are only allowed in the (degenerate) {\it end} components, and precisely two of them in each such end. 
See Fig.~\ref{fig:TypeIIIab} for an illustration of this refinement. 

\begin{figure}[h]
  \center
  \includegraphics[width=0.45\textwidth]{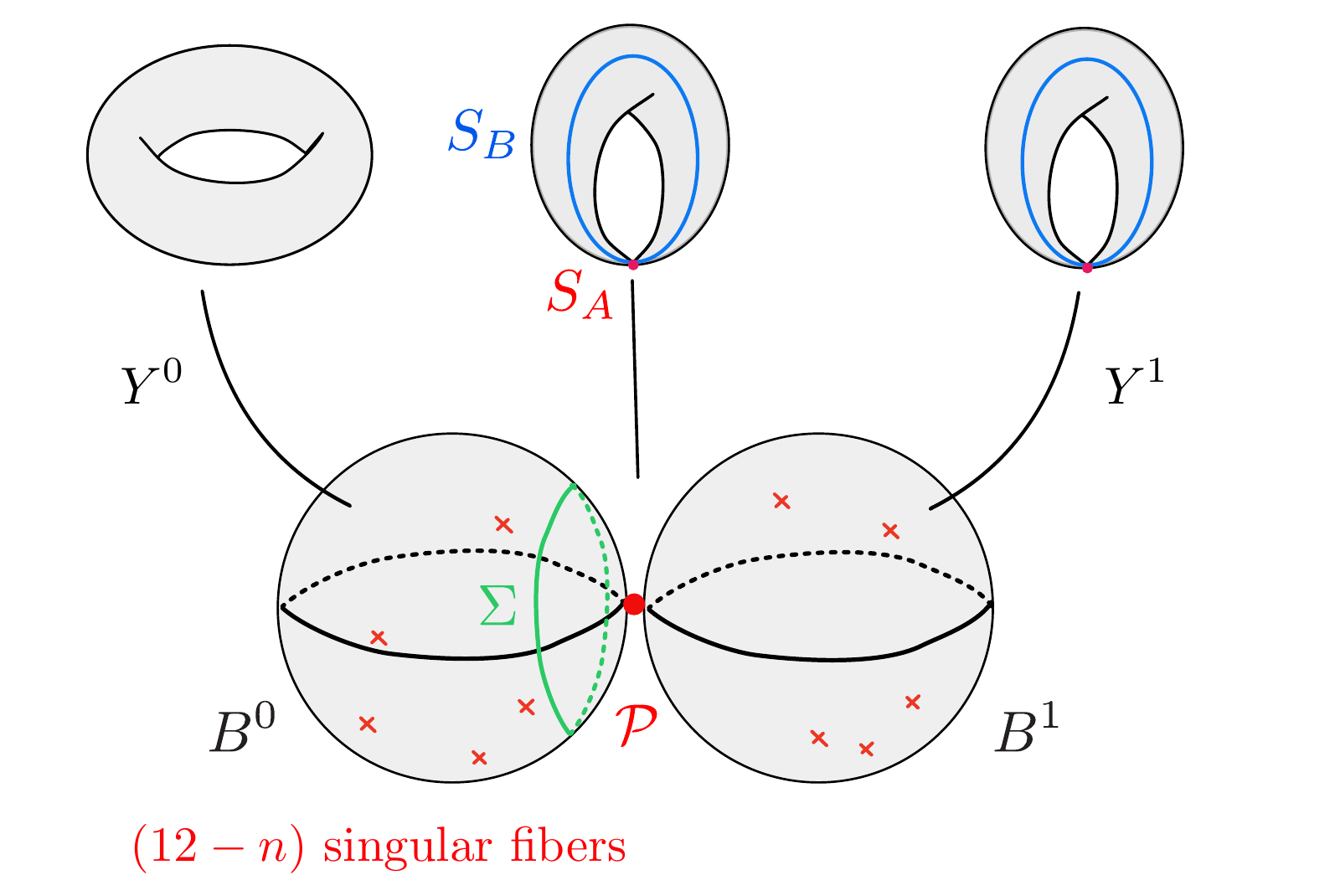}
  \caption{\it The characteristic topology of a Type III.a Kulikov Weierstrass model at its rational elliptic end component, say, $Y_0$, adjoining the degenerate component $Y^1$ with generic I$_{n}$ fibers. Red crosses on the base components denote the locations of singular fibers, $12-n$ of which sit on $B^0$.}
  \label{fig:TypeIIIa}
\end{figure}

As it turns out, the effective physics associated with the two subtypes of elliptic Type III models parallel their Type II counterparts to some extent and exhibit important distinguishing features at the same time. F-theory on a Type III.a model is again dual to heterotic string on a torus, but with both the K\"ahler and the complex structure parameters taken to infinity ($T_{\rm het} \sim U_{\rm het} \to \infty$), and hence, the spacetime partially decompactifies to $9d$. Similarly, a Type III.b model also renders the Type IIB string weakly-coupled, but on top of that decompactifies the spacetime to $10d$. Such phases of the effective physics are easily read off from the geometric structure as follows~\cite{Lee:2021usk}. 
\bi
\item For models of  Type III.a, decompactification of one circular direction is indicated in M-theory by the light KK-like tower, which results from M2 branes wrapping the shrinking transcendental 2-torus $\gamma_1$. The latter in turn arises from fibering the A-cycle of the elliptic fiber along the vanishing 1-cycle encircling the base point $B^p\cap B^{p+1}$ for any $p$;\footnote{The generic singular fibers of Kodaira type $I_{n_p}$ turn out to be mutually local in a global sense~\cite{Lee:2021usk} so that a common 2-torus class $\gamma_1$ is defined for all $p$. Furthermore, the B-cycle cannot be fibered along the base 1-cycle due to the $I_n$ monodromy around the point $B^p \cap B^{p+1}$.} see Fig.~\ref{fig:TypeIIIa} for a schematic description of the intersection with $p=0$.
These states correspond in F-theory to the tower resulting from $(1,0)$-strings wound around the same vanishing 1-cycle of the base. 
\item For models of Type III.b, there arises a tensionless $(1,0)$-string from M2 brane wrapping the fibral A-cycle, which vanishes {\it globally} over the entire base $B_0$ unlike in models of Type III.a. However, two (or more) D-type fibers, and hence also the associated O-planes, are bound to collide in $\hat B_0$ of the degenerate K3 surface~\eqref{hatY0}, since otherwise the base blowups would not have occurred to begin with. This indicates a complex structure degeneration of the Type IIB torus and hence, in turn, a decompactification of the spacetime to $10d$. Note that the 2-torus $\gamma_1$ is defined in the same way as in Type III.a case and leads to a light KK-like tower once wrapped by M2 branes. The other light tower, however, needs to be inferred indirectly. 
\ei

Let us close this section by providing an alternative F-theoretic interpretation of the light towers resulting from Type III.a degenerations, with the central notion being the affine extension,\footnote{A similar interpretation can be given to Type II.a models, where the involved extension is of rank-$2$: $E_8 \to \hat E_8 \to \hat E_9$.}
\beq\label{aff}
E_{9-n}   \longrightarrow   \hat E_{9-n} \,, \quad  n \geq 1\,,
\eeq
associated with the $SL(2, \IZ)$ monodromy action of a brane stack formed at infinite distance. 
Specifically, as depicted in Fig.~\ref{fig:TypeIIIa}, at the intersection of a rational elliptic end component, say, $Y^0$, and its adjacent degenerate component $Y^1$ lies an I$_{n}$ fiber, where $n:=n_1 \geq 1$. 
Since the combined monodromy action of the $12$ singular fibers on $B^0$ should be trivial, we immediately learn that a total of $12-n$ branes on $B^0$ away from $\cP := B^0 \cap B^1$ generate the affine algebra $\hat E_{9-n}$, of which monodromy action is the inverse of the monodromy of $A_{n-1}$. In the described affinization~\eqref{aff} occurring at infinite distance,\footnote{A comprehensive analysis of monodromy actions was given in~\cite{DeWolfe:1998zf}, which is suggestive of potential formation of $\hat E_{9-n}$ brane stacks. Kulikov models of Type III.a explicitly realize such affine extensions at infinite distance in the complex structure moduli.} the imaginary root leads to a BPS winding tower via $(1,0)$-string. This thus serves as a light KK tower, indicating that the spacetime partially decompactifies to $9d$. 

Apparently, the imaginary root only provides a KK tower and is ignored in the gauge algebra of the decompactified theory. We thus propose that the $9d$ gauge algebra acquires an $E_{9-n}$ factor from the branes on $B^0$. As an immediate consequence of this interpretation, we learn that the maximal gauge algebras\footnote{Strictly speaking, these may only constitute the non-abelian part of the gauge algebra. However, the claimed algebras already saturate the rank bound $17$. We thus believe that additional $u(1)$ factors cannot be present, which is natural also in that all of the $24$ branes have already been used up.} in ``$9d$ F-theory'' are $A_{m+n-1} \oplus E_{9-n} \oplus E_{9-m}$ (if both end components are rational elliptic) or $D_{8+n} \oplus E_{9-n}$ (if only one end is rational elliptic), where the constraints~\eqref{AD} on the allowed brane types have been used. Such gauge algebras have indeed been constructed in terms of explicit Kulikov Weierstrass models at infinite distance, practically by engineering appropriate non-minimal fibers in codimension one~\cite{Lee:2021usk}. It is assuring that our F-theoretic results are in complete agreement with the recent classification from the heterotic side~\cite{Font:2020rsk},\footnote{See also the previous analysis of~\cite{Cachazo:2000ey}.} where the $9d$ algebras were analyzed via Wilson lines on $S^1$. This non-trivial match serves as an independent piece of evidence for our decompactification proposal.

\section{Conclusions and Prospects}\label{Conc}
In this article we have reported on recent progress towards identifying towers of states expected to become light at infinite distance in moduli space of quantum gravity. Specifically, in the context of F-theory compactifications, the physical nature of such light towers has been clarified either as the excitation modes of an emergent weakly-coupled string or as the Kaluza-Klein modes associated with a decompactification, in agreement with the Emergent String Conjecture. 

We started by geometrically classifying the infinite distance limits, both in the K\"ahler and in the complex structure moduli spaces of F-theory, respectively, in $4$ and $8$ dimensions. We have then proceeded to investigate the effective theories resulting from such asymptotic forms of the compactification geometry, and have thus confirmed, to begin with, that a tower of states does become light. This calls for an alternative way of describing the effective physics in the boundary of F-theory moduli space, notably, in terms of perturbative {\it heterotic} or {\it Type II} strings. The addressed light tower of states thus clarifies precisely how one should switch between different duality frames of string theory. 

Both in the K\"ahler and in the complex structure sides of the story, we have seen that a pivotal role is played by the vanishing of non-contractible cycles, which we identified via our geometric classification. In all infinite distance limits, those shrinking cycles, once wrapped by various appropriate branes in the theory, lead either to a light KK-like tower of particles or to an asymptotically tensionless critical string, which in turn generates a light tower of excitation particles. In the latter situation, the geometry can in principle allow for multiple different species of lightest critical strings, which may look pathological at first sight. However, we have seen that in all such cases certain cycles in the geometry expand fast in such a way that the associated KK scale is parametrically lighter than the string scale, clarifying the fate of the limit as a decompactification phase instead. 

In principle, effective tensionless branes with more than one spatial extensions may also arise in the same vein, again as a solitonic object in the effective theory~\cite{Font:2019cxq}. In many cases, however, their tensions do not lead to a lightest scale of the effective theory. Furthermore, in other cases where classical geometry alone does allow for higher-dimensional branes of which tension determines the lightest scale, quantum corrections in the moduli space turn out to obstruct those classical limits~\cite{Alvarez-Garcia:2021pxo}, thereby supporting the Emergent String Conjecture. This is similar in spirit to how quantum corrections obstruct seemingly pathological limits where the KK scale is lighter than the string scale, as discussed in Section~\ref{Kahler}. It remains to be seen if and how analogous quantum effects may arise for infinite complex structure limits.  

Admittedly, our analysis in this article has been strictly in the context of string theory. The appearance of light towers has nonetheless proven to be by far non-trivial and is thus suggestive of a potential effective-theoretic interpretation of the Emergent String Conjecture.\footnote{See e.g.~\cite{Lanza:2020qmt, Lanza:2021udy} for an EFT string point of view on infinite distance limits.} Progress along this line would even more strengthen our understanding of light towers at infinite distance and could possibly serve as strong evidence of string theory as {\it the} framework of quantum gravity.

\section*{Acknowledgements}
The author is grateful to Daniel Klaewer, Wolfgang Lerche, Guglielmo Lockhart, Timo Weigand and Max Wiesner for collaborations on the projects which this article is based upon. 
This work is supported by IBS under the project code, IBS-R018-D1.

\bibliography{papers}

\providecommand{\href}[2]{#2}\begingroup\raggedright\begin{thebibliography}{10}

\bibitem{Vafa:2005ui}
C.~Vafa, {\it {The String landscape and the swampland}},
  \href{http://arxiv.org/abs/hep-th/0509212}{{\tt hep-th/0509212}}.

\bibitem{Ooguri:2006in}
H.~Ooguri and C.~Vafa, {\it {On the Geometry of the String Landscape and the
  Swampland}},  {\em Nucl. Phys. B} {\bf 766} (2007) 21--33,
  [\href{http://arxiv.org/abs/hep-th/0605264}{{\tt hep-th/0605264}}].

\bibitem{Brennan:2017rbf}
T.~D. Brennan, F.~Carta, and C.~Vafa, {\it {The String Landscape, the
  Swampland, and the Missing Corner}},  {\em PoS} {\bf TASI2017} (2017) 015,
  [\href{http://arxiv.org/abs/1711.00864}{{\tt arXiv:1711.00864}}].

\bibitem{Palti:2019pca}
E.~Palti, {\it {The Swampland: Introduction and Review}},  {\em Fortsch. Phys.}
  {\bf 67} (2019), no.~6 1900037, [\href{http://arxiv.org/abs/1903.06239}{{\tt
  arXiv:1903.06239}}].

\bibitem{vanBeest:2021lhn}
M.~van Beest, J.~Calder\'on-Infante, D.~Mirfendereski, and I.~Valenzuela, {\it
  {Lectures on the Swampland Program in String Compactifications}},
  \href{http://arxiv.org/abs/2102.01111}{{\tt arXiv:2102.01111}}.

\bibitem{Grana:2021zvf}
M.~Gra\~na and A.~Herr\'aez, {\it {The Swampland Conjectures: A Bridge from
  Quantum Gravity to Particle Physics}},  {\em Universe} {\bf 7} (2021), no.~8
  273, [\href{http://arxiv.org/abs/2107.00087}{{\tt arXiv:2107.00087}}].

\bibitem{Lee:2019wij}
S.-J. Lee, W.~Lerche, and T.~Weigand, {\it {Emergent Strings from Infinite
  Distance Limits}},  \href{http://arxiv.org/abs/1910.01135}{{\tt
  arXiv:1910.01135}}.

\bibitem{Vafa:1996xn}
C.~Vafa, {\it {Evidence for F theory}},  {\em Nucl. Phys. B} {\bf 469} (1996)
  403--418, [\href{http://arxiv.org/abs/hep-th/9602022}{{\tt hep-th/9602022}}].

\bibitem{Corvilain:2018lgw}
P.~Corvilain, T.~W. Grimm, and I.~Valenzuela, {\it {The Swampland Distance
  Conjecture for K\"ahler moduli}},  {\em JHEP} {\bf 08} (2019) 075,
  [\href{http://arxiv.org/abs/1812.07548}{{\tt arXiv:1812.07548}}].

\bibitem{Baume:2019sry}
F.~Baume, F.~Marchesano, and M.~Wiesner, {\it {Instanton Corrections and
  Emergent Strings}},  {\em JHEP} {\bf 04} (2020) 174,
  [\href{http://arxiv.org/abs/1912.02218}{{\tt arXiv:1912.02218}}].

\bibitem{Blumenhagen:2018nts}
R.~Blumenhagen, D.~Kl\"awer, L.~Schlechter, and F.~Wolf, {\it {The Refined
  Swampland Distance Conjecture in Calabi-Yau Moduli Spaces}},  {\em JHEP} {\bf
  06} (2018) 052, [\href{http://arxiv.org/abs/1803.04989}{{\tt
  arXiv:1803.04989}}].

\bibitem{Grimm:2018ohb}
T.~W. Grimm, E.~Palti, and I.~Valenzuela, {\it {Infinite Distances in Field
  Space and Massless Towers of States}},  {\em JHEP} {\bf 08} (2018) 143,
  [\href{http://arxiv.org/abs/1802.08264}{{\tt arXiv:1802.08264}}].

\bibitem{Joshi:2019nzi}
A.~Joshi and A.~Klemm, {\it {Swampland Distance Conjecture for One-Parameter
  Calabi-Yau Threefolds}},  {\em JHEP} {\bf 08} (2019) 086,
  [\href{http://arxiv.org/abs/1903.00596}{{\tt arXiv:1903.00596}}].

\bibitem{Grimm:2019ixq}
T.~W. Grimm, C.~Li, and I.~Valenzuela, {\it {Asymptotic Flux Compactifications
  and the Swampland}},  {\em JHEP} {\bf 06} (2020) 009,
  [\href{http://arxiv.org/abs/1910.09549}{{\tt arXiv:1910.09549}}]. [Erratum:
  JHEP 01, 007 (2021)].

\bibitem{Gendler:2020dfp}
N.~Gendler and I.~Valenzuela, {\it {Merging the weak gravity and distance
  conjectures using BPS extremal black holes}},  {\em JHEP} {\bf 01} (2021)
  176, [\href{http://arxiv.org/abs/2004.10768}{{\tt arXiv:2004.10768}}].

\bibitem{Grimm:2020cda}
T.~W. Grimm, {\it {Moduli space holography and the finiteness of flux vacua}},
  {\em JHEP} {\bf 10} (2021) 153, [\href{http://arxiv.org/abs/2010.15838}{{\tt
  arXiv:2010.15838}}].

\bibitem{Grimm:2021ikg}
T.~W. Grimm, J.~Monnee, and D.~van~de Heisteeg, {\it {Bulk Reconstruction in
  Moduli Space Holography}},  \href{http://arxiv.org/abs/2103.12746}{{\tt
  arXiv:2103.12746}}.

\bibitem{Bastian:2021eom}
B.~Bastian, T.~W. Grimm, and D.~van~de Heisteeg, {\it {Modeling General
  Asymptotic Calabi-Yau Periods}},  \href{http://arxiv.org/abs/2105.02232}{{\tt
  arXiv:2105.02232}}.

\bibitem{Palti:2021ubp}
E.~Palti, {\it {Stability of BPS States and Weak Coupling Limits}},
  \href{http://arxiv.org/abs/2107.01539}{{\tt arXiv:2107.01539}}.

\bibitem{Bastian:2021hpc}
B.~Bastian, T.~W. Grimm, and D.~van~de Heisteeg, {\it {Engineering Small Flux
  Superpotentials and Mass Hierarchies}},
  \href{http://arxiv.org/abs/2108.11962}{{\tt arXiv:2108.11962}}.

\bibitem{Grimm:2021ckh}
T.~W. Grimm, E.~Plauschinn, and D.~van~de Heisteeg, {\it {Moduli Stabilization
  in Asymptotic Flux Compactifications}},
  \href{http://arxiv.org/abs/2110.05511}{{\tt arXiv:2110.05511}}.

\bibitem{Xu:2020nlh}
F.~Xu, {\it {On TCS G$_{2}$ manifolds and 4D emergent strings}},  {\em JHEP}
  {\bf 10} (2020) 045, [\href{http://arxiv.org/abs/2006.02350}{{\tt
  arXiv:2006.02350}}].

\bibitem{Lee:2018urn}
S.-J. Lee, W.~Lerche, and T.~Weigand, {\it {Tensionless Strings and the Weak
  Gravity Conjecture}},  {\em JHEP} {\bf 10} (2018) 164,
  [\href{http://arxiv.org/abs/1808.05958}{{\tt arXiv:1808.05958}}].

\bibitem{Lee:2018spm}
S.-J. Lee, W.~Lerche, and T.~Weigand, {\it {A Stringy Test of the Scalar Weak
  Gravity Conjecture}},  {\em Nucl. Phys. B} {\bf 938} (2019) 321--350,
  [\href{http://arxiv.org/abs/1810.05169}{{\tt arXiv:1810.05169}}].

\bibitem{Lee:2019tst}
S.-J. Lee, W.~Lerche, and T.~Weigand, {\it {Modular Fluxes, Elliptic Genera,
  and Weak Gravity Conjectures in Four Dimensions}},  {\em JHEP} {\bf 08}
  (2019) 104, [\href{http://arxiv.org/abs/1901.08065}{{\tt arXiv:1901.08065}}].

\bibitem{Klaewer:2020lfg}
D.~Klaewer, S.-J. Lee, T.~Weigand, and M.~Wiesner, {\it {Quantum corrections in
  4d $N$ = 1 infinite distance limits and the weak gravity conjecture}},  {\em
  JHEP} {\bf 03} (2021) 252, [\href{http://arxiv.org/abs/2011.00024}{{\tt
  arXiv:2011.00024}}].

\bibitem{Arkani-Hamed:2006emk}
N.~Arkani-Hamed, L.~Motl, A.~Nicolis, and C.~Vafa, {\it {The String landscape,
  black holes and gravity as the weakest force}},  {\em JHEP} {\bf 06} (2007)
  060, [\href{http://arxiv.org/abs/hep-th/0601001}{{\tt hep-th/0601001}}].

\bibitem{Cheung:2014vva}
C.~Cheung and G.~N. Remmen, {\it {Naturalness and the Weak Gravity
  Conjecture}},  {\em Phys. Rev. Lett.} {\bf 113} (2014) 051601,
  [\href{http://arxiv.org/abs/1402.2287}{{\tt arXiv:1402.2287}}].

\bibitem{Heidenreich:2015nta}
B.~Heidenreich, M.~Reece, and T.~Rudelius, {\it {Sharpening the Weak Gravity
  Conjecture with Dimensional Reduction}},  {\em JHEP} {\bf 02} (2016) 140,
  [\href{http://arxiv.org/abs/1509.06374}{{\tt arXiv:1509.06374}}].

\bibitem{Heidenreich:2016aqi}
B.~Heidenreich, M.~Reece, and T.~Rudelius, {\it {Evidence for a sublattice weak
  gravity conjecture}},  {\em JHEP} {\bf 08} (2017) 025,
  [\href{http://arxiv.org/abs/1606.08437}{{\tt arXiv:1606.08437}}].

\bibitem{Montero:2016tif}
M.~Montero, G.~Shiu, and P.~Soler, {\it {The Weak Gravity Conjecture in three
  dimensions}},  {\em JHEP} {\bf 10} (2016) 159,
  [\href{http://arxiv.org/abs/1606.08438}{{\tt arXiv:1606.08438}}].

\bibitem{Lee:2020gvu}
S.-J. Lee, W.~Lerche, G.~Lockhart, and T.~Weigand, {\it {Quasi-Jacobi forms,
  elliptic genera and strings in four dimensions}},  {\em JHEP} {\bf 01} (2021)
  162, [\href{http://arxiv.org/abs/2005.10837}{{\tt arXiv:2005.10837}}].

\bibitem{Lee:2020blx}
S.-J. Lee, W.~Lerche, G.~Lockhart, and T.~Weigand, {\it {Holomorphic Anomalies,
  Fourfolds and Fluxes}},  \href{http://arxiv.org/abs/2012.00766}{{\tt
  arXiv:2012.00766}}.

\bibitem{Palti:2017elp}
E.~Palti, {\it {The Weak Gravity Conjecture and Scalar Fields}},  {\em JHEP}
  {\bf 08} (2017) 034, [\href{http://arxiv.org/abs/1705.04328}{{\tt
  arXiv:1705.04328}}].

\bibitem{Kulikov1}
V.~Kulikov, {\it {Degenerations of K3 surfaces and Enriques surfaces}},  {\em
  Math. USSR Izv. 11} {\bf 05} (1977) 957--989.

\bibitem{Kulikov2}
V.~Kulikov, {\it {On modifications of degenerations of surfaces with $\kappa =
  0$}},  {\em Math. USSR Izv. 17} {\bf 02} (1981) 339--342.

\bibitem{PerssonPink}
U.~Persson and H.~Pinkham, {\it {Degenerations of surfaces with trivial
  canonical bundle}},  {\em Ann of Math. (2) 113} {\bf 01} (1981) 45--66.

\bibitem{FriedmanMorrison}
R.~Friedman and D.~Morrison, {\it {The birational geometry of degenerations: An
  overview}},  {\em Progr Math.} {\bf 29} (1983) 1--32.

\bibitem{Persson}
U.~Persson, {\it {On degenerations of algebraic surfaces}},  {\em Mem. Amer.
  Math. Soc. 11} {\bf 189} (1977).

\bibitem{Clingher:2003ui}
A.~Clingher and J.~W. Morgan, {\it {Mathematics underlying the F theory /
  Heterotic string duality in eight-dimensions}},  {\em Commun. Math. Phys.}
  {\bf 254} (2005) 513--563, [\href{http://arxiv.org/abs/math/0308106}{{\tt
  math/0308106}}].

\bibitem{Morrison:1996na}
D.~R. Morrison and C.~Vafa, {\it {Compactifications of F theory on Calabi-Yau
  threefolds. 1}},  {\em Nucl. Phys.} {\bf B473} (1996) 74--92,
  [\href{http://arxiv.org/abs/hep-th/9602114}{{\tt hep-th/9602114}}].

\bibitem{Morrison:1996pp}
D.~R. Morrison and C.~Vafa, {\it {Compactifications of F theory on Calabi-Yau
  threefolds. 2.}},  {\em Nucl. Phys.} {\bf B476} (1996) 437--469,
  [\href{http://arxiv.org/abs/hep-th/9603161}{{\tt hep-th/9603161}}].

\bibitem{Aspinwall:1997ye}
P.~S. Aspinwall and D.~R. Morrison, {\it {Point - like instantons on K3
  orbifolds}},  {\em Nucl. Phys. B} {\bf 503} (1997) 533--564,
  [\href{http://arxiv.org/abs/hep-th/9705104}{{\tt hep-th/9705104}}].

\bibitem{Sen:1996vd}
A.~Sen, {\it {F theory and orientifolds}},  {\em Nucl. Phys. B} {\bf 475}
  (1996) 562--578, [\href{http://arxiv.org/abs/hep-th/9605150}{{\tt
  hep-th/9605150}}].

\bibitem{Clingher:2012rg}
A.~Clingher, R.~Donagi, and M.~Wijnholt, {\it {The Sen Limit}},  {\em Adv.
  Theor. Math. Phys.} {\bf 18} (2014), no.~3 613--658,
  [\href{http://arxiv.org/abs/1212.4505}{{\tt arXiv:1212.4505}}].

\bibitem{Lee:2021qkx}
S.-J. Lee and T.~Weigand, {\it {Elliptic K3 Surfaces at Infinite Complex
  Structure and their Refined Kulikov models}},
  \href{http://arxiv.org/abs/2112.07682}{{\tt arXiv:2112.07682}}.

\bibitem{alexeev2021compactifications}
V.~Alexeev, A.~Brunyate, and P.~Engel, {\it Compactifications of moduli of
  elliptic k3 surfaces: stable pair and toroidal},
  \href{http://arxiv.org/abs/2002.07127}{{\tt arXiv:2002.07127}}.

\bibitem{Brunyantethesis}
A.~Brunyate, {\it A modular compactification of the space of elliptic k3
  surfaces},  {\em PhD Thesis at University of Georgia} (2015).

\bibitem{Odaka:2018kua}
Y.~Odaka and Y.~Oshima, {\it {Collapsing K3 surfaces, Tropical geometry and
  Moduli compactifications of Satake, Morgan-Shalen type}},
  \href{http://arxiv.org/abs/1810.07685}{{\tt arXiv:1810.07685}}.

\bibitem{Odaka:2020ezm}
Y.~Odaka, {\it {PL density invariant for type II degenerating K3 surfaces,
  Moduli compactification and hyperKahler metrics}},
  \href{http://arxiv.org/abs/2010.00416}{{\tt arXiv:2010.00416}}.

\bibitem{ascher2021compact}
K.~Ascher and D.~Bejleri, {\it Compact moduli of elliptic k3 surfaces},  2021.

\bibitem{Lee:2021usk}
S.-J. Lee, W.~Lerche, and T.~Weigand, {\it {Physics of Infinite Complex
  Structure Limits in eight Dimensions}},
  \href{http://arxiv.org/abs/2112.08385}{{\tt arXiv:2112.08385}}.

\bibitem{DeWolfe:1998zf}
O.~DeWolfe and B.~Zwiebach, {\it {String junctions for arbitrary Lie algebra
  representations}},  {\em Nucl. Phys. B} {\bf 541} (1999) 509--565,
  [\href{http://arxiv.org/abs/hep-th/9804210}{{\tt hep-th/9804210}}].

\bibitem{Font:2020rsk}
A.~Font, B.~Fraiman, M.~Gra\~na, C.~A. N\'u\~nez, and H.~P. De~Freitas, {\it
  {Exploring the landscape of heterotic strings on $T^d$}},  {\em JHEP} {\bf
  10} (2020) 194, [\href{http://arxiv.org/abs/2007.10358}{{\tt
  arXiv:2007.10358}}].

\bibitem{Cachazo:2000ey}
F.~A. Cachazo and C.~Vafa, {\it {Type I' and real algebraic geometry}},
  \href{http://arxiv.org/abs/hep-th/0001029}{{\tt hep-th/0001029}}.

\bibitem{Font:2019cxq}
A.~Font, A.~Herr\'aez, and L.~E. Ib\'a\~nez, {\it {The Swampland Distance
  Conjecture and Towers of Tensionless Branes}},  {\em JHEP} {\bf 08} (2019)
  044, [\href{http://arxiv.org/abs/1904.05379}{{\tt arXiv:1904.05379}}].

\bibitem{Alvarez-Garcia:2021pxo}
R.~\'Alvarez-Garc\'\i{}a, D.~Kl\"awer, and T.~Weigand, {\it {Membrane Limits in
  Quantum Gravity}},  \href{http://arxiv.org/abs/2112.09136}{{\tt
  arXiv:2112.09136}}.

\bibitem{Lanza:2020qmt}
S.~Lanza, F.~Marchesano, L.~Martucci, and I.~Valenzuela, {\it {Swampland
  Conjectures for Strings and Membranes}},  {\em JHEP} {\bf 02} (2021) 006,
  [\href{http://arxiv.org/abs/2006.15154}{{\tt arXiv:2006.15154}}].

\bibitem{Lanza:2021udy}
S.~Lanza, F.~Marchesano, L.~Martucci, and I.~Valenzuela, {\it {The EFT stringy
  viewpoint on large distances}},  {\em JHEP} {\bf 09} (2021) 197,
  [\href{http://arxiv.org/abs/2104.05726}{{\tt arXiv:2104.05726}}].

\end{thebibliography}\endgroup
\bibliographystyle{JHEP}

\end{document}